\documentstyle[12pt,amssymb]{article}
\begin{document}
\tolerance=5000
\def\pp{{\, \mid \hskip -1.5mm =}}
\def\cL{{\cal L}}
\def\be{\begin{equation}}
\def\ee{\end{equation}}
\def\bea{\begin{eqnarray}}
\def\eea{\end{eqnarray}}
\def\tr{{\rm tr}\, }
\def\nn{\nonumber \\}
\def\e{{\rm e}}

\begin{titlepage}

\ \hfill OCHA-PP-223

\begin{center}
\Large
{\bf Freedom in electroweak symmetry breaking and mass matrix of fermions in 
dimensional deconstruction model}

\vfill

\normalsize

Shin'ichi Nojiri$^\spadesuit$\footnote{Electronic mail: nojiri@nda.ac.jp, 
snojiri@yukawa.kyoto-u.ac.jp}, 
Sergei D. Odintsov$^{\heartsuit\clubsuit}$\footnote{Electronic mail: 
odintsov@ieec.fcr.es Also at TSPU, Tomsk, Russia}, 

and Akio Sugamoto$^\diamondsuit$\footnote{Electronic mail: sugamoto@phys.ocha.ac.jp}

\normalsize

\vfill

{\em $\spadesuit$ Department of Applied Physics, 
National Defence Academy, \\
Hashirimizu Yokosuka 239-8686, JAPAN}

\ 

{\em $\heartsuit$ Institut d'Estudis Espacials de Catalunya (IEEC), \\
Edifici Nexus, Gran Capit\`a 2-4, 08034 Barcelona, SPAIN}

\ 

{\em $\clubsuit$ Instituci\`o Catalana de Recerca i Estudis 
Avan\c{c}ats (ICREA), \\Barcelona, SPAIN}

\ 

{\em $\diamondsuit$ Dept. of Physics, Faculty of Science, Ochanomizu Univ. \\
1-1 Otsuka 2, Bunkyo-ku, Tokyo 112-8610, JAPAN
}

\end{center}

\vfill 

\baselineskip=24pt
\begin{abstract}

There exists a freedom in a class of four-dimensional electroweak theories 
proposed by Arkani-Hamed et al. relying on deconstruction and Coleman-Weinberg 
mechanism.  The freedom comes from the winding modes of the link variable 
(Wilson operator) connecting non-nearest neighbours in the discrete fifth 
dimension. Using this freedom, dynamical breaking of SU(2) gauge symmetry, 
mass hierarchy patterns of fermions and Cabbibo-Kobayashi-Maskawa matrix may be 
obtained.

\end{abstract}


\end{titlepage}

Dimensional deconstruction \cite{AH-C-G, AH} (see also subsequent 
works \cite{AH2, others, Sugamoto, AH3, 
AH4, AH5, Falkowski, KS}) suggests the natural electroweak (EW) symmetry 
breaking in four dimensions without using supersymmetry or strong dynamics 
at the TeV scale physics. 
The interesting feature of such approach is that perturbative corrections 
in Higgs sector are finite. 
In these models, the extra dimensions are the discrete lattice. 
The simplest version \cite{AH-C-G} is given by the sites on a circle. 
By the Coleman-Weinberg mechanism \cite{CW}, the gauge symmetry can be broken 
spontaneously. The effective potential of the Higgs field 
becomes finite. In the naive model for $SU(2)$ gauge theory, the Higgs field 
is, however, triplet, that is, in the adjoint representation. In the realistic 
models, of course, the Higgs field should be an $SU(2)$ doublet. In order to 
introduce the doublet Higgs field, the $N\times N$ torus (moose) of the lattice has been 
introduced \cite{AH-C-G} and has been investigated in \cite{AH3}. Especially the 
simplest case of $N=2$ torus case has been constructed in \cite{AH4} and the most 
economical case that the Higgs field is pseudo Goldstone boson in an $SU(5)/SO(5)$ 
has been presented in \cite{AH5}.  

However, it is quite important (which was not properly realized
so far) that there may be  more freedom in EW symmetry 
breaking patterns from dimensional deconstruction due to its dependence 
from the non-nearest neighbour couplings in theory space. 
Eventually, it means that bigger number of phenomenologically accepted EW
symmetry breakings may be
realized by the corresponding choice of boundary condition from latticized
dimension. The variety of novel Higgs sectors may emerge.
In the context of the $U(1)^N$ model, 
Wilson-line operators with arbitrary couplings have been discussed in \cite{Falkowski}, 
where such operators are generated with finite coefficients by radiative corrections. 
The non-nearest neighbour couplings in theory space, however, may appear geometrically. 
We may assume the sites on the lattice corresponding to the branes embedded in the higher 
dimensional spacetime. Then the link variables connecting the different sites may 
correspond to the open string. Hence if the embedding space has non-trivial homotopy 
($S^1$ is most simple but non-trivial case), there will appear several couplings 
corresponding to the winding mode of the open strings (say, an open string can connect 
two different branes after winding $S^1$ several times). When the embedding manifold $M$ 
is compact but the dimensions (codimension of the brane) is larger than one, the homotopy 
$\pi_1(M)$ can be more complicated and the non-nearest neighbour couplings appear in general. 
Another interpretation for such couplings could be that continuum limit of 
higher dimensional gauge theory is non-local. 
Thus, we consider a (non-linear) generalization of the simplest
model where the sites of 
the lattice lies on a circle. Then the Higgs fields are in the adjoint representation. 
Therefore the theory under consideration is a non-linear, toy model for EW
symmetry
breaking.
More realistic models probably may be constructed by considering the
discrete torus as in 
\cite{AH2,AH3,AH4}. 


Here we propose a generalization of the model \cite{AH-C-G}, so that it 
may qualitatively describe the dynamical EW breaking and the mass
hierarchy of quarks or leptons between different generations\footnote{
The model with an infinite number of gauge theories which are linked by 
scalars has been considered in \cite{HS} in order to get an infinite 
tower of massive gauge fields. The model  \cite{HS} may be in a same 
class with that in \cite{AH-C-G}. We also note that a generalization 
of the model 
 \cite{AH-C-G} by using the graph structure has been done in \cite{KS}.}.
The model  \cite{AH-C-G} includes $N$-copies of the gauge field $A_\mu^n$ and
$N$ link variables $U_{n,n+1}$, following \cite {AH}. For the link variables, we 
impose a periodic boundary condition $U_{n+N,n+N+1}=U_{n,n+1}$ and sometimes 
we restrict $n$ to be $n=0,1,2,\cdots,N-1$.  $U_{n,n+1}$ is assumed to be unitary 
and  $U_{n+1,n}$ is defined by $U_{n+1,n}\equiv U_{n,n+1}^\dagger =
U_{n,n+1}^{-1}$. 

Before writing our new Lagrangian, it is convenient to define a variable
$U_{n,l}$, a link variable connecting ``the non-nearest neighbours'', by
\be
\label{E1}
U_{n,l}=\left\{\begin{array}{ll}
U_{n,n+1}U_{n+1,n+2}\cdots U_{l-2,l-1}U_{l-1,l}\quad & \mbox{when}\ l>n \\
1 & \mbox{when}\ l=n \\
U_{n,n-1}U_{n-1,n-2}\cdots U_{l+2,l+1}U_{l+1,l}\quad & \mbox{when}\ l<n \ .\\
\end{array}\right. 
\ee
The following Lagrangian is the main starting point of our new model:
\be
\label{E2}
{\cal L}=-{1 \over 2g^2}\sum_{n=0}^{N-1} \tr F^n_{\mu\nu}F^{n\,\mu\nu} 
+ {1 \over 4}\sum_{n,l}a_{nl}\tr \left[\left(D_\mu U_{n,l}\right)^\dagger D^\mu 
U_{n,l}\right]\ .
\ee
Here $F^n_{\mu\nu}$ is the field strength given by $A_\mu^n$ and $a_{nl}$'s are 
constants 
specifying the couplings including non-nearest neighbours. This kind of 
couplings was first discussed for gravity in \cite{E}, and is useful to
obtain 
the induced positive cosmological constant, which may serve 
as a quite simple model for the dark energy of our accelerating universe. 
In the model \cite{AH-C-G}, only  nearest neighbour couplings have been
introduced. 
If we assume $U_{n,l}$ connects the branes, in the present model, the branes are connected 
in a rather complicated way. One may suppose the branes correspond to the
site on a circle. 
Then $U_{n,l}$ connects the branes like a mesh or a net. Such a case might not 
occur if the codimension of the spacetime is one. We may need to consider more 
complicated spacetime or the spacetime whose codimension is two or more.  

If we denote the gauge group as $G$, the Lagrangian (\ref{E2}) has $G^N$ gauge 
symmetry.  
The non-nearest neighbour couplings in (\ref{E2}) give more degrees of
freedom to the model, and are useful to trigger the dynamical breaking of 
gauge symmetry and the mass hierarchy. 
As it will be shown later, the induced Coleman-Weinberg potential and the mass 
matrix of fermions can
include an 
arbitrary function originating from the non-nearest neighbour couplings.  
The proper choice of the functions induces the gauge symmetry breaking and the 
mass hierarchy.

Since $U_{n,l}\neq U_{n,l+N}$ nor $U_{n,l}\neq U_{n+N,l}$ in general, the sums 
about $n$ and 
$l$ can be from $-\infty$ to $\infty$. 
In (\ref{E2}), the covariant derivative $D_\mu$ is defined by
\be
\label{E3}
D_\mu U_{n,l}\equiv \partial_\mu U_{n,l} - i A_\mu^n U_{n,l} + i U_{n,l} A_\mu^l\ .
\ee
In the Lagrangian (\ref{E2}), the terms ${\cal L}_M$ which do not include 
derivative 
can be regarded as mass terms for the gauge fields and ${\cal L}_M$ is 
explicitly given 
by
\be
\label{E4}
{\cal L}_M = {1 \over 4}\sum_{n,l}a_{nl}\tr \left[A_\mu^n A^{n\,\mu} 
+ A^l_\mu A^{l\,\mu} - 2 A^l_\mu U_{l,n} A^{n\,\mu} U_{n,l}\right]\ .
\ee
By using the gauge transformation, one may impose the unitary gauge
condition where 
$U_{n,n+1}$ does not depend on $n$ as
\be
\label{E5}
U_{n,n+1}=\e^{iu}\ .
\ee

First we consider the electrodynamics case, where the gauge group is $U(1)$. 
Since $U_{l,n}^\dagger=U_{n,l}$, and $U_{l,n}$ commutes with $A^{n\,\mu}$, 
one obtains 
\be
\label{EM1}
{\cal L}_M = {1/2}\sum_{n,l}a_{nl} \left[A_\mu^{n} A^{n\,\mu} 
+ A^{l}_\mu A^{l\,\mu} -2 A_\mu^{n} A^{l\,\mu}\right]\ .
\ee
There does not appear $u$-dependence.

As a non-trivial toy model,  the case that the gauge group is $SU(2)$ is
interesting. 
Then the action (\ref{E2}) has $SU(2)^N$ gauge symmetry.
Writing
\be
\label{E7}
A^n_\mu = {1 \over 2}\tau^a A^{na}_\mu\ ,\quad u={\upsilon_0 \over 2}\tau^3\ .
\ee
with $\tau^a$'s $\left(a=1,2,3\right)$ being Pauli matrices, we find
\bea
\label{E8}
{\cal L}_M &=& {1 \over 2}\sum_{n,l}a_{nl} \left[A_\mu^{na} A^{na\,\mu} 
+ A^{la}_\mu A^{la\,\mu} \right.\nn
&& - 2 \cos\left(\left(l-n\right)\upsilon_0\right)\left(
A_\mu^{n1} A^{l1\,\mu}+A_\mu^{n2} A^{l2\,\mu}\right) \nn
&& \left. + 2 \sin\left(\left(l-n\right)\upsilon_0\right)\left(
A_\mu^{l1} A^{n2\,\mu} - A_\mu^{l2} A^{n1\,\mu}\right)
-2 A_\mu^{n3} A^{l3\,\mu}\right]\ .
\eea
We may assume $a_{nl}$ only depends on the absolute value of the difference 
between 
$n$ and $l$:
\be
\label{E6}
a_{nl}=a\left(\left|n-l\right|\right)\ .
\ee
We also Fourier transform $A^{na}_\mu$ as 
\be
\label{E9}
A^{na}_\mu={1 \over \sqrt{N}}\sum_{L=0}^{N-1}\hat A^{La}_\mu \e^{i{2\pi nL 
\over N}}\ .
\ee
Since $A^{na}_\mu$ is real,  $\hat A^{LA}_\mu = \left(\hat 
A^{(N-L)A}_\mu\right)^*$. 
Then we obtain 
\bea
\label{E10}
{\cal L}_M &=& 2\sum_{L=0}^{N-1}\sum_{l=1}^\infty a(l) \left[
\left(1 - \cos\left(l\upsilon_0\right)\cos\left({2\pi lL \over N}\right)\right) 
\right.\nn
&& \times \left( \left(\hat A_\mu^{L1}\right)^* \hat A^{L1\,\mu} 
+ \left(\hat A_\mu^{L2}\right)^* \hat A^{L2\,\mu}\right) \nn
&& +  \sin\left(l\upsilon_0\right)\sin\left({2\pi lL \over N}\right) \left(
\left(\hat A_\mu^{L1}\right)^* \hat A^{L2\,\mu} 
 - \left(\hat A_\mu^{L2}\right)^* \hat A^{L1\,\mu}\right) \nn
&& \left. + \left(1 - \cos\left({2\pi lL \over N}\right)\right) 
\left(\hat A_\mu^{L3}\right)^* \hat A^{L3\,\mu}\right]\ .
\eea
As a result the mass matrix for the gauge fields is given by 
\bea
\label{E11}
&& M_n^2\left(\upsilon_0\right)= 32g^2 \left(\begin{array}{ccc}
{1 \over \sqrt{2}} & {i \over \sqrt{2}} & 0 \\
{i \over \sqrt{2}} & {1 \over \sqrt{2}} & 0 \\
0 & 0 & 1 
\end{array}\right) \nn
&& \times \left\{\sum_{l=1}^\infty a(l) 
\left(\begin{array}{ccc}
2\sin^2\left(l\left({\upsilon_0 \over 2} + {\pi L \over N}\right)\right) & 0 & 
0 \\
0 & 2\sin^2\left(l\left({\upsilon_0 \over 2} - {\pi L \over N}\right)\right) & 
0 \\
0 & 0 & 2\sin^2\left( {\pi lL \over N}\right) 
\end{array}\right)\right\} \nn
&& \times \left(\begin{array}{ccc}
{1 \over \sqrt{2}} & -{i \over \sqrt{2}} & 0 \\
-{i \over \sqrt{2}} & {1 \over \sqrt{2}} & 0 \\
0 & 0 & 1 
\end{array}\right)\ .
\eea
When $\upsilon_0=0$, only the mode corresponding to $L=0$ is
massless. Then $SU(2)^N$ gauge 
symmetry is broken to $SU(2)$. When $\upsilon_0\neq 0$, the massless gauge 
field is only 
$\hat A_\mu^{L=0,a=3}$. Thus, the gauge symmetry is broken down to $U(1)$. 

In order to obtain non-vanishing $\upsilon_0$, one may consider the
Coleman-Weinberg 
mechanism \cite{CW}, where the one-loop induced potential 
for $\upsilon_0$ is given by
\be
\label{E14}
V\left(\upsilon_0\right) = {3\Lambda^2 \over 32\pi^2} \tr \left( 
M_n^2\left(\upsilon_0\right) \right) 
+ {3 \over 64\pi^2} \tr\left\{ \left(M_n^2\left(\upsilon_0\right)\right)^2 \ln 
\left( 
{M_n^2\left(\upsilon_0\right) \over \Lambda^2}\right)\right\}\ .
\ee
Here $\Lambda^2$ is the UV cut-off parameter. 
As the kinetic term for $\upsilon_0$ in (\ref{E2}) is given by 
\be
\label{E12}
{\cal L}_K = {1 \over 4}\sum_{l=1}^\infty a(l) l^2 \partial_\mu \upsilon_0 
\partial^\mu \upsilon_0 \ ,
\ee 
the canonically normalized field $\phi$ is  
\be
\label{E13}
\phi = \upsilon_0 \sqrt{\sum_{l=1}^\infty a(l) l^2 \over 2} \ .
\ee

It is interesting to consider now some examples. Let $f(x)$ be a function
which can be expanded by a Taylor 
series:
\be
\label{E15}
f(x)=\sum_{k=0}^\infty \alpha_k x^k\ .
\ee
 $a(l)$ is chosen as
\be
\label{E16}
a(l)=\left\{\begin{array}{ll}
\alpha_k \ & \mbox{when}\ l=Nk+1\ (k=0,1,2,\cdots) \\
0 & \mbox{when}\ l\neq Nk+1 
\end{array}\right. \ .
\ee
Here $k$ can be regarded as the winding number. 
As a result
\bea
\label{E17}
&& 2\sum_{l=1}^\infty a(l) 
\sin^2\left(l\left({\upsilon_0 \over 2} \pm {\pi L \over N}\right)\right) \nn
&& = f(1) - {1 \over 2}\left\{
\e^{i\left(\upsilon_0 \pm {2\pi L \over N}\right)} 
f\left(\e^{iN\upsilon_0}\right)
+ \e^{-i\left(\upsilon_0 \pm {2\pi L \over N}\right)} 
f\left(\e^{-iN\upsilon_0}\right)\right\}\ .
\eea
Therefore we have 
\be
\label{E18}
2 \sum_{L=0}^{N-1} \sum_{l=1}^\infty a(l) 
\sin^2\left(l\left({\upsilon_0 \over 2} \pm {\pi L \over N}\right)\right) = N 
f(1)\ .
\ee
Then in the potential (\ref{E14}), the term proportional to $\Lambda^2$ does 
not 
depend on $\upsilon_0$. 
We also have 
\bea
\label{E19}
&& \sum_{L=0}^{N-1} \left(2\sum_{l=1}^\infty a(l) 
\sin^2\left(l\left({\upsilon_0 \over 2} \pm {\pi L \over 
N}\right)\right)\right)^2 \nn
&& = \left\{\begin{array}{ll}
N\left\{f(1)^2 + {1 \over 
2}f\left(\e^{iN\upsilon_0}\right)f\left(\e^{-iN\upsilon_0}\right)\right\}\ 
& \mbox{when}\ N>2 \\
2f(1)^2 + {1 \over 2}\left\{\e^{i \upsilon_0 } f\left(\e^{2i\upsilon_0}\right)
+ \e^{-i\upsilon_0} f\left(\e^{-2i\upsilon_0}\right)\right\}^2 \ & \mbox{when}\ 
N=2 
\end{array}\right. \ .
\eea
When $N>2$, if $f(x)\propto x^I$ by a non-negative integer $I$, 
in the potential (\ref{E14}), the term proportional to $\ln \Lambda^2$ does not 
depend on $\upsilon_0$. The model with $f(x)\propto x^I$ does not have any
essential 
difference with that in \cite{AH-C-G}, since $f(x)$ is a monomial means not to 
include different winding modes. 
On the other hand, in the general case in which $f(x)$ is a polynomial including 
different winding modes, this term depends on $\upsilon_0$. 
In the model \cite{AH-C-G}, there does not appear $\ln \Lambda^2$ terms
in the 
field (corresponding to $\upsilon_0$ here) dependent part. 
In the potential we now have included $\ln \Lambda^2$ term in general but
in return for it, 
we have a degrees of freedom of an arbitrary (Taylor expansible) function 
$f(x)$. 
As  $U_{n,l}$ with $|n-l|>1$ is included, even for the case of $N=2$, the
potential can be 
rather different from that in \cite{AH-C-G}. 
The potential (\ref{E14}) for $N=2$ case is explicitly found to be
\bea
\label{E20}
V\left(\upsilon_0\right) &=& 
{96 g^4 \over \pi^2} \left[\left\{ f(1)^2 + {1 \over 4}\left\{\e^{i \upsilon_0 
} f\left(\e^{2i\upsilon_0}\right)
+ \e^{-i\upsilon_0} f\left(\e^{-2i\upsilon_0}\right)\right\}^2 \right\}\right. 
\nn
&& \times \ln \left\{ {f(1)^2 - {1 \over 4 }\left\{\e^{i \upsilon_0 } 
f\left(\e^{2i\upsilon_0}\right)
+ \e^{-i\upsilon_0} f\left(\e^{-2i\upsilon_0}\right)\right\}^2 \over \Lambda^4} 
\right\}  \nn
&& - f(1) \left\{\e^{i \upsilon_0 } f\left(\e^{2i\upsilon_0}\right)
+ \e^{-i\upsilon_0} f\left(\e^{-2i\upsilon_0}\right)\right\} \nn
&& \left. \times \ln \left\{ {f(1) - {1 \over 2}\left\{\e^{i \upsilon_0 } 
f\left(\e^{2i\upsilon_0}\right)
+ \e^{-i\upsilon_0} f\left(\e^{-2i\upsilon_0}\right)\right\} \over 
f(1) + {1 \over 2}\left\{\e^{i \upsilon_0 } f\left(\e^{2i\upsilon_0}\right)
+ \e^{-i\upsilon_0} f\left(\e^{-2i\upsilon_0}\right)\right\}}\right\}\right] \nn
&& + \left(\mbox{$\upsilon_0$ independent terms}\right) \ .
\eea
If we define
\be
\label{X1}
X^\pm \equiv f(1) \pm {1 \over 2}\left\{\e^{i \upsilon_0 } f\left(\e^{2i\upsilon_0}\right)
+ \e^{-i\upsilon_0} f\left(\e^{-2i\upsilon_0}\right)\right\}\ ,
\ee
$V\left(\upsilon_0\right)$ in (\ref{E20}) can be rewritten as 
\bea
\label{X2}
V\left(\upsilon_0\right) &=& 
{96 g^4 \over \pi^2} \left[ {{X^+}^2 + {X^-}^2 \over 2}\ln \left({X^+ X^- \over \Lambda^4}
\right) - {{X^+}^2 - {X^-}^2 \over 2}\ln \left({X^+ \over X^-}\right)\right] \nn
&& + \left(\mbox{$\upsilon_0$ independent terms}\right) \nn
&=&\left. {96 g^4 \over \pi^2} \right[{{X^-}^2 \over 2}\ln \left(X^-\right)^2 
+ {\left(2f(1) - X^-\right)^2 \over 2}\ln \left(2f(1) - X^-\right)^2 \nn
&& \left. - {\left(2f(1) - X^-\right)^2 + {X^-}^2 \over 2}\ln \Lambda^4\right] \nn
&& + \left(\mbox{$\upsilon_0$ independent terms}\right)\ .
\eea
We should note that $\upsilon_0=0$ corresponds to $X^-=0$. 
When $\upsilon_0$ is small, we find
\be
\label{X2b}
X^-\sim \left({1 \over 2}f(1) + 4f'(1) -2f''(1)\right)\upsilon_0^2\ .
\ee
When $\upsilon_0$ ($X^-$) is small, 
the potential $V\left(\upsilon_0\right)$ behaves as
\bea
\label{X3}
\lefteqn{V\left(\upsilon_0\right)=\left.
{96 g^4 \over \pi^2} \right[ \left(\mbox{$\upsilon_0$ independent terms}\right)
} \nn
&& \left.+ \left( - 4f(1)\ln \left(2f(1)\right) - 2 + 2f(1) \ln \Lambda^4 \right)X^- 
+ {\cal O}\left({X^-}^2 \ln X^-\right)\right] \nn
&\sim& \left. {96 g^4 \over \pi^2} \right[ 
\left(\mbox{$\upsilon_0$ independent terms}\right) \nn
&& + \left( - 4f(1)\ln \left(2f(1)\right) - 2 + 2f(1) \ln \Lambda^4 \right) \nn
&& \left.\times \left({1 \over 2}f(1) + 4f'(1) -2f''(1)\right)\upsilon_0^2 \right] \ .
\eea
As the term linear in $X^-$ appears, the point $X^-=0$ ($\upsilon_0=0$) is unstable 
in general. For example, for the choice $f(x)=1 - {x \over 2}$, one gets
\be
\label{X3b}
V\left(\upsilon_0\right)\sim {96 g^4 \over \pi^2} \left[ 
\left(\mbox{$\upsilon_0$ independent terms}\right)
 - {3 \over 2} \left( - 2 + \ln \Lambda^4 \right)\upsilon_0^2\right] \ .
\ee
If $\Lambda^4 >\e^2$, the coefficient of $\upsilon_0^2$ becomes negative and 
the point of $\upsilon_0=0$ is unstable and there could be a non-trivial 
vacuum expectation value. 
Note that with the choice $f(x)=1 - {x \over 2}$, $X^\pm$ are given by
\be
\label{X3c}
X^\pm = \mp 2\left(\cos\upsilon_0 \pm 1 \right) 
\left(\cos^2\upsilon_0 \mp \cos\upsilon_0 - {1 \over 4}\right)\ .
\ee
Since $\left|\cos\upsilon_0\right|\leq 1$, $X^-$ is bounded as
\be
\label{X3d}
{1 \over 2}\left\{-\left({5 \over 3}\right)^{3 \over 2} + 1\right\}\leq X^-
\leq {1 \over 2}\left\{\left({5 \over 3}\right)^{3 \over 2} + 1\right\}\ .
\ee
$X^-$ has maximum at $\cos\upsilon_0=-{1 \over 2}\sqrt{5 \over 3}$ 
and  minimum at $\cos\upsilon_0={1 \over 2}\sqrt{5 \over 3}$.
In the region given by (\ref{X3d}), $V\left(\upsilon_0\right)$  (\ref{X2}) is finite, 
then $V\left(\upsilon_0\right)$ is bounded and has finite maximum and minimum.
Eq.(\ref{X3b}) tells that at the minimum, $\upsilon_0$ does not vanish.

One sees
\bea
\label{E21}
{1 \over 2}\sum_{l=1}^\infty a(l) l^2 
&=& {1 \over 2} \sum_{k=0}^\infty \alpha_k \left(Nk + 1\right)^2 \nn
&=& {1 \over 2}\left\{ N^2\left(f''(1) + f'(1)\right) + 2Nf'(1) + f(1) \right\}
\eea
for general $N$. Then for the case of $N=2$, the canonically normalized field 
$\phi$ 
in (\ref{E13}) is given by
\be
\label{E22}
\phi = \upsilon_0 \sqrt{
4 f''(1) + 8 f'(1) + f(1) 
 \over 2} \ .
\ee

In the similar way the coupling with the spinor fields which may be
identified with quarks or leptons could be introduced. 
In order to specify the theory, one may restrict the gauge symmetry to be
$SU(2)$ and 
 the spinors of 2-dimensional representation of $SU(2)$ may be considered:
\be
\label{E23}
\Psi^n=\left(\begin{array}{c} u^n \\ d^n \end{array}\right)\ .
\ee
If we regard the spinors as quarks, we may identify $u^n$ as up-type quarks and 
$d^n$ as down-types ones. Then a rather general Lagrangian with general $N$ has 
the following form:
\be
\label{E24}
{\cal L}_f =i \sum_{n=0}^{N-1}\bar \Psi^n D_\mu \left(A_\mu^n\right)
\gamma^\mu \Psi^n - \sum_{n=0}^{N-1} \sum_{l=-\infty}^\infty b_{nl} \bar\Psi^n 
U_{n,l} \Psi^l \ .
\ee
Here $\Psi^{n+N}=\Psi^n$.
First we assume as in $a_{nl}$ of (\ref{E6}) that $b_{nl}$ depends on the 
absolute value 
of $n-l$:
\be
\label{E25}
b_{nl}=b\left(\left|n-l\right|\right)\ .
\ee
Choosing $U_{n,n+1}$ as in (\ref{E5}) with (\ref{E7}), one finds
\bea
\label{E26}
&& \sum_{n=0}^{N-1} \sum_{l=-\infty}^\infty b_{nl} \bar\Psi^n U_{n,l} \Psi^l \nn
&& = \sum_{L=0}^{N-1}\left(\bar{\hat u^L}, \bar{\hat d^L}\right)
\left(\begin{array}{cc} B(L) & 0 \\ 0 & B(L) \end{array}\right) 
\left(\begin{array}{c} \hat u^L \\ \hat d^L \end{array}\right)\ .
\eea
Here we have written $\Psi^n$ as 
\be
\label{E27}
\Psi^n = {1 \over \sqrt{N}}\sum_{L=0}^{N-1}\hat\Psi^L \e^{i{2\pi nL \over N}} 
= {1 \over \sqrt{N}}\sum_{L=0}^{N-1}\left(\begin{array}{c} 
\hat u^L \\ \hat d^L \end{array}\right) \e^{i{2\pi nL \over N}} \ ,
\ee
and $B(L)$ is defined by
\be
\label{E28}
B(L)=b(0) + 2\sum_{l=1}^\infty \cos \left(l\left({\upsilon_0 \over 2} 
+ {2\pi L \over N}\right)\right) b(l)\ .
\ee
Let $g(x)$ is an arbitrary even function which can be expanded as a 
Fourier series
\be
\label{E29}
g(x)=\sum_{l=-\infty}^\infty g_l \e^{ilx} 
= g_0 + 2 \sum_{l=1}^\infty g_l \cos\left(lx\right)\ ,
\ee
and the identification is done:
\be
\label{E30}
b(l)=g_l\ .
\ee
Then 
\be
\label{E31}
B(L)=g\left({\upsilon_0 \over 2} + {2\pi L \over N}\right)\ .
\ee
In order to specify the model we consider $N=3$ case.
Choosing $g(x)$ 
as an exponential function, for example
\be
\label{E32}
g=\zeta \e^{\eta x}
\ee
with $\zeta\e{\eta \upsilon_0 \over 2} \sim 10$ MeV and $\eta \sim 2$, we find 
$B(0)\sim 10$ MeV, $B(1)\sim 1$ GeV, $B(2)\sim 100$ GeV. Hence, we may
identify 
the quark of $L=0$ as $(u,d)$, $L=1$ as $(c,s)$, and $L=2$ as $(b,t)$. 

Although we may explain the hierarchy between the generation of the quarks by 
the mass 
matrix (\ref{E26}), the mass of the up-type quarks and that of the down-type ones 
are 
degenerate in (\ref{E26}) and there does not appear Cabbibo-Kobayashi-Maskawa 
matrix. 
In order to solve the problem,  the assumption in (\ref{E25}) may be
discarded. 
Now we introduce $N$ arbitrary real functions $g_{nn}(x)$ $n=0,\cdots N-1$ 
and ${N(N-1) \over 2}$ arbitrary complex functions $g_{nm}(x)$ 
$\left(N-1 \geq n>m\geq 0\right)$.  
The function $g_{nm}(x)$ with $\left(n<m\right)$ are defined by a complex 
conjugate of
$g_{nm}(x)$: 
$g_{nm}(x)\equiv g_{mn}(x)^*$. The natural assumption is that these
functions can be expanded as a 
Fourier series: 
\be
\label{E33}  
g_{nm} (x)=\sum_{k=-\infty}^\infty g_{nm}^k\e^{ikx} \ .
\ee
With identification $b_{nl}$ by 
\be
\label{E34}
b_{n\, m+Nk} = g_{nm}^k\ .
\ee
one gets 
\bea
\label{E35}
&& \sum_{n=0}^{N-1} \sum_{l=-\infty}^\infty b_{nl}\bar\Psi^n U_{n,l} \Psi^l =
\sum_{n,m=0}^{N-1} \bar\Psi^n \tilde U_{n,m} \Psi^m \nn
&& = \sum_{n,m=0}^{N-1}\left(\bar u^n, \bar d^n\right)
\left(\begin{array}{cc} G_{nm}\left(\upsilon_0\right) & 0 
\\ 0 & G_{nm}\left(-\upsilon_0\right) \end{array}\right) 
\left(\begin{array}{c} u^n \\ d^m \end{array}\right)\ .
\eea
Here 
\be
\label{SS1}
\tilde{U}_{nm}\equiv \sum_{k=-\infty}^\infty b_{n\, m+kN} U_{n,m+kN}\ ,
\ee
and 
\be
\label{E36}
G_{nm}\left(\upsilon_0\right) \equiv \e^{i{(m-n)\upsilon_0 \over 
2}}g_{nm}\left({N\upsilon_0 \over 2}\right)
\ee
We should note that the loop of the fermion gives an additional contribution 
to the Coleman-Weinberg 
potential in (\ref{E14}) by 
\bea
\label{E14f}
V_f\left(\upsilon_0\right) &=& 
 -{\Lambda^2 \over 32\pi^2} \tr \left( \tilde U \left(\upsilon_0\right)\tilde 
U^\dagger\left(\upsilon_0\right)\right) \nn
&& - {1 \over 64\pi^2} \tr\left\{ \left(\tilde U \left(\upsilon_0\right)\tilde 
U^\dagger\left(\upsilon_0\right)\right)^2 \ln \left( 
{\tilde U \left(\upsilon_0\right)\tilde U^\dagger\left(\upsilon_0\right) 
\over \Lambda^2}\right)\right\}\ .
\eea
If we regard $G_{nm}\left(\upsilon_0\right)$ as an $N\times N$ matrix, 
$G_{nm}\left(\upsilon_0\right)$ is 
hermitian: $G_{nm}\left(\upsilon_0\right)=G_{mn}\left(\upsilon_0\right)^*$.
Hence, we can diagonalize 
$G_{nm}\left(\upsilon_0\right)$ by an $N\times N$ unitary matrix 
$V_{nm}\left(\upsilon_0\right)$:
\be
\label{E37}
\sum_{n',m'=0}^{N-1}V_{nn'}\left(\upsilon_0\right) 
G_{n'm'}\left(\upsilon_0\right)
V_{m'm}\left(\upsilon_0\right)^\dagger = G_n 
\left(\upsilon_0\right)\delta_{nm}\ .
\ee
Then the mass eigenstates in (\ref{E35}) are given by
\be
\label{E38}
\tilde u^n = \sum_{n'=0}^{N-1} V_{nn'}\left(\upsilon_0\right)u^{n'}\ ,\quad
\tilde d^n = \sum_{n'=0}^{N-1} V_{nn'}\left(-\upsilon_0\right)d^{n'} \ .
\ee
The mass eigenvalues of up-type quarks are given by 
$G_n\left(\upsilon_0\right)$. 
On the other hand, the mass eigenvalues of down-type quarks are given by
$G_n\left(-\upsilon_0\right)$. 
Since $G_n\left(-\upsilon_0\right)\neq G_n\left(\upsilon_0\right)$ in general, 
the masses of up-type quarks 
can be different from those of down-type quarks. The gauge couplings in the 
Lagrangian 
(\ref{E24}) are:
\be
\label{E39}
\bar u \left(\gamma^\mu A_\mu\right) d 
= \sum_{n,m,n'=0}^{N-1}\bar{\tilde u^n} \left(\gamma^\mu A_\mu\right) 
V_{nn'}\left(\upsilon_0\right)V_{n'm}\left(-\upsilon_0\right)^\dagger \tilde 
d^m\ .
\ee
Since $\tilde u^n$ and $\tilde d^m$ are the eigenstates of the mass, we may 
identify 
$M_{nm}\equiv \sum_{n'=0}^{N-1} 
V_{nn'}\left(\upsilon_0\right)V_{n'm}\left(-\upsilon_0\right)^\dagger$ 
as the Cabbibo-Kobayashi-Maskawa (CKM) matrix. 

In this paper we have given a generalization of the 5 dimensional model 
studied by Arkani-Hamed et al.\cite{AH-C-G} using $N$ branes and $N$ copies of 
fields and 
symmetries \cite{AH}.  The new point implemented 
in our paper is the introduction of 
the link variables $U_{n,l}$ which connect branes ($n$-th and $l$-th) in the 
non-nearest neighborhood. Since the link variables in the 5th dimension 
give the Higgs fields, our model becomes a new model of the Higgs sector. 
If the 5th dimension is considered as a discrete circle made from $N$ points, 
possibly, our link variables have winding numbers with respect to the 
discrete circle.  Owing to this non-nearest neighbour link variables, we 
have obtained the following interesting results:
\begin{enumerate}
\item The dynamical breaking of gauge symmetry occurs, and
\item the quark (or lepton) masses and the CKM mixing matrix are dynamically 
induced, reproducing the mass hierarchy.
\end{enumerate}

More explicitly, in a model with $SU(2)$ gauge symmetry, the set of 
coefficients $\{\alpha_k\}$ in (\ref{E16}) ($k$: winding number) of the kinetic 
terms for 
$U_{n, n+1+Nk}$ gives a function $f(x)$ which determines the non-vanishing 
vacuum expectation value of the Higgs scalar, following the 
Coleman-Weinberg mechanism.  Accordingly the $SU(2)$ gauge symmetry is 
dynamically broken down to $U(1)$.  This phenomenon occurs even for $N=2$, but 
does not occur without incorporating the different winding sectors.

Regarding the fermion mass matrix, we have studied the $SU(2)$ model.  Here 
the total number $N$ of branes becomes the number of generations of quarks or 
leptons, and the different waves of fermion fields standing on the discrete 
$N$ points give the different generations. Therefore, the model with $N=3$ is a 
three generation model. The $SU(2)$ symmetry used here is not $SU(2)_{L}$ nor 
$SU(2)_{R}$, but the diagonal group of $SU(2)_{V}$.  The coefficients 
$b_{mn}^{k}$ of $\bar{\Psi} U_{m, n+Nk} \Psi$ correspond to the Yukawa 
couplings. Here the Yukawa couplings also have the winding number $k$ with 
respect to the discrete circle of $N$ points.  Similarly as before, the 
winding number dependence of the Yukawa couplings gives a function 
$g_{mn}(x)$ which  determines the mass eigenvalues and CKM mixing matrix of 
the model.

In a simplified case with $SU(2)$ symmetry, the masses $m_1, m_2, m_3, ...,$ 
and $m_{N}$ of 1st, 2nd, 3rd, ..., and $N$-th generation fermions, 
respectively, give the hierarchical structure following
\bea
\label{S1}
&& m_1:m_2:m_3: \cdots:m_{N} \nn
&& =g\left({\upsilon_0 \over 2}+{2\pi \over N}\right):
g\left({\upsilon_0 \over 2} +{4\pi \over N}\right):
g\left({\upsilon_0 \over 2}+ {6\pi \over N}\right): \nn
&& \qquad \cdots: g\left({\upsilon_0 \over 2}+2\pi\right).
\eea
where $\upsilon_0$ is the vacuum expectation value of the Higgs scalar 
determined 
dynamically \` a la Coleman and 
Weinberg, and $N$ is the number of generations.
Since the $SU(2)$ symmetry can be broken dynamically, the mass matrix of up-type 
quarks, and that of down-type quarks can be determined differently, giving 
different hierarchical structure.

As was stated above we have used the $SU(2)_{V}$ symmetry, and  the 
left-right asymmetry is not incorporated, so that the mixing matrices of 
L-handed current (CC interaction) and the R-handed current (not yet observed) 
are identical. 
This result itself is not bad in our study focused on the Higgs sector.  We 
have to eliminate, however, the R-handed current from our model and 
construct a realistic $SU(2) \times U(1)$ model in order to obtain a 
realistic gauge sector.  For this purpose, we have to introduce the 
chirality of branes on which the fermions with the same chirality live. 
Then, the gauge interaction comes from the connection of the branes with 
the same chiralities and the Higgs interactions connect those with the 
opposite chiralities (see ref.\cite{Sugamoto}).

In summary, we should stress that the toy, non-linear deconstruction model 
discussed above is not realistic in the same way as first model 
\cite{AH-C-G}. Moreover, some properties of such
non-linear theory  (like UV completion, its continuum interpretation,
etc) are not quite clear and should be further investigated. Nevertheless,
in our
opinion, the
additional freedom which the new deconstruction model
 may introduce to EM symmetry breaking patterns may be useful in the
generalization of more  realistic versions\cite{AH3,AH4,AH5}.

\ 

\noindent{\bf Acknowledgments}

We thank M. Quiros for helpful discussions.
This investigation has been supported in part  by the Ministry of
Education, Science, Sports and Culture of Japan under the grant n.13135208
(S.N.), n. 1403920 (A.S.), and by DGICYT (Spain), project BFM2003-00620 (S.D.O.).


\begin{thebibliography}{99}
\bibitem{AH-C-G} N. Arkani-Hamed, A.G. Cohen and H. Georgi, {\sl Phys.Lett.} 
{\bf B513} (2001) 232, hep-ph/0105239.
\bibitem{AH} N. Arkani-Hamed, A.G. Cohen and H. Georgi, {\sl Phys.Rev.Lett.} {\bf 86} 
(2001) 4757-4761, hep-th/0104005.
\bibitem{AH2} N. Arkani-Hamed, H. Georgi and M.D. Schwartz, {\sl Ann.Phys.NY} {\bf 305} 
(2003) 96-118, hep-th/0210184; \\
C.T. Hill, S. Pokorski and J. Wang, {\sl Phys.Rev.} {\bf D64} (2001) 105005.
\bibitem{others} M. Bander, {\sl Phys.Rev.} {\bf D64} (2001) 105021;
V. Jejjala, R. Leigh and D. Minic, {\sl Phys.Lett.} {\bf B556} (2003) 71. 
\bibitem{Sugamoto} A. Sugamoto, {\sl Prog.Theor.Phys.} {\bf 107} (2002) 793, hep-th/0104241; \\
A. Sugamoto, {\sl Grav.Cosmol.} {\bf 9} (2003) 91, hep-th/0210235.
\bibitem{AH3} N. Arkani-Hamed, A.G. Cohen, T. Gregoire and J.G. Wacker, 
{\sl JHEP} {\bf 0208} (2002) 020, hep-ph/0202089. 
\bibitem{AH4} N. Arkani-Hamed, A.G. Cohen, E. Katz, A.E. Nelson, 
T. Gregoire and J.G. Wacker, {\sl JHEP} {\bf 0208} (2002) 021,
hep-ph/0206020.
\bibitem{AH5} N. Arkani-Hamed, A.G. Cohen, E. Katz and A.E. Nelson, {\sl
JHEP} 
{\bf 0207} (2002) 034, hep-ph/0206021.
\bibitem{Falkowski} A. Falkowski, C. Grojean and S. Pokorski, {\sl
Phys.Lett.} 
{\bf B581} (2004) 236, hep-ph/0310201. 



\bibitem{KS} N. Kan and K. Shiraishi, gr-qc/0310055. 

\bibitem{CW} S. Coleman and E. Weinberg, {\sl Phys.Rev.} {\bf D7}
(1973) 1888.  




\bibitem{HS} M.B. Halpern, W. Siegel, {\sl Phys.Rev.} {\bf D11} (1975) 2967. 


\bibitem{E} G. Cognola, E. Elizalde, S. Nojiri, S.D. Odintsov and S. Zerbini,
hep-th/0312269.
\end{thebibliography}
\end{document}